\documentclass[sigconf,dvipsnames]{acmart}
\renewcommand\footnotetextcopyrightpermission[1]{}
\acmConference[EASE 2025]{The 29th International Conference on Evaluation and Assessment in Software Engineering}{17–20 June, 2025}{Istanbul, Türkiye}

\usepackage{lscape}
\usepackage{graphicx}
\usepackage{hyperref}
\usepackage{titlesec}
\usepackage{tcolorbox}
\usepackage{enumitem}
\usepackage{balance}
\usepackage{soul}
\usepackage{algorithmic}
\usepackage{listings}
\usepackage{comment}

\AtBeginDocument{%
  \providecommand\BibTeX{{%
    Bib\TeX}}}

\setcopyright{acmlicensed}
\copyrightyear{2018}
\acmYear{2018}
\acmDOI{XXXXXXX.XXXXXXX}

\newcommand{\lesson}[1]{\vspace{0.5em}\setlength{\fboxsep}{0.015\linewidth}\noindent\fcolorbox{BlueViolet}{white}{\parbox{0.96\linewidth}{\textcolor{BlueViolet}{} #1}} \vspace{0.5em}}

\def\BibTeX{{\rm B\kern-.05em{\sc i\kern-.025em b}\kern-.08em
    T\kern-.1667em\lower.7ex\hbox{E}\kern-.125emX}}

\begin{document}

\begin{tcolorbox}[colback=yellow!10!white, colframe=yellow!50!black]
\centering \textbf{This is a preprint version. The final version will appear in the proceedings of EASE 2025.}
\end{tcolorbox}

\title{On Simulation-Guided LLM-based Code Generation\\
for Safe Autonomous Driving Software
}

\author{Ali Nouri $^{1,2}$, Johan Andersson $^{2}$, Kailash De Jesus Hornig $^{2}$, Zhennan Fei$^{1, 2}$\\ Emil Knabe $^{1}$, Håkan Sivencrona $^{1}$, Beatriz Cabrero-Daniel $^{2,3}$, Christian Berger $^{2,3}$\\[0.5ex]
$^{1}$\textit{Volvo Cars, Gothenburg, Sweden}\\
$^{2}$\textit{Chalmers University of Technology, Department of Computer Science and Engineering},\\
$^{3}$\textit{University of Gothenburg, Department of Computer Science and Engineering},\\
\{ali.nouri, zhennan.fei, hakan.sivencrona\} @volvocars.com,\\
\{beatriz.cabrero-daniel, christian.berger\} @gu.se
}

\renewcommand{\shortauthors}{Nouri et al.}
\renewcommand{\shorttitle}{On Simulation-Guided LLM-based Code Generation for Safe Autonomous Driving Software}




\begin{abstract}
Automated Driving System (ADS) is a safety-critical software system responsible for the interpretation of the vehicle's environment and making decisions accordingly. The unbounded complexity of the driving context, including unforeseeable events, necessitate continuous improvement, often achieved through iterative DevOps processes. However, DevOps processes are themselves complex, making these improvements both time- and resource-intensive. Automation in code generation for ADS using Large Language Models (LLM) is one potential approach to address this challenge. Nevertheless, the development of ADS requires rigorous processes to verify, validate, assess, and qualify the code before it can be deployed in the vehicle and used.
In this study, we developed and evaluated a prototype for automatic code generation and assessment using a designed pipeline of a LLM-based agent, simulation model, and rule-based feedback generator in an industrial setup. The LLM-generated code is evaluated automatically in a simulation model against multiple critical traffic scenarios, and an assessment report is provided as feedback to the LLM for modification or bug fixing.
We report about the experimental results of the prototype employing Codellama:34b, DeepSeek (r1:32b and Coder:33b), CodeGemma:7b, Mistral:7b, and GPT4 for Adaptive Cruise Control (ACC) and Unsupervised Collision Avoidance by Evasive Manoeuvre (CAEM).
We finally assessed the tool with 11 experts at two Original Equipment Manufacturers (OEMs) by conducting an interview study.
\end{abstract}



\received{31 January 2025}
\received[accepted]{31 March 2025}

\maketitle

\section{Introduction}

Autonomous Driving Systems (ADS) use various onboard sensors and vehicle-to-everything (V2X) communication to perceive the environment to decide and control a vehicle's steering, acceleration, and braking. This enables autonomous navigation on roads, reducing the need for human intervention.
However, ensuring ADS safety is challenging due to the unbounded complexity of the Operational Design Domain (ODD) and the fact that ADS itself is a software-intensive system of systems.

One approach to addressing the challenge is to limit the functionality (e.g., automatic lane keeping) and its ODD to specific situations (e.g., good visibility in daylight), although the ultimate goal is to cover all driving scenarios to meet SAE Level 5 capabilities~\cite{SAEJ3016_202104}.
The function is closely monitored during operation to identify unknown hazardous scenarios~\cite{sotif} and rapidly implement necessary improvements to prevent future incidents.
As more confidence is gained through continuous monitoring, the functionality and its ODD are gradually expanded.
To facilitate the continuous expansion of ADS functionality and its ODD, software-defined vehicles~\cite{24safeopsbosch} rely heavily on software, while the hardware remains largely unchanged. Over-the-air (OTA) updates and centralized compute units enable the ongoing enhancement of functionalities during operation, leveraging DevOps~\cite{Nouri2022SEAA}.
Rapid software updates are not only crucial for guaranteeing the safety of the system against new, unknown hazardous situations but also for improving customer experience by continuously expanding functionality. Automation is an enabler for increasing speed and improving efficiency in design, development, and verification activities.

Natural language serves as the main input in various stages of the software engineering process, including function descriptions, requirements engineering~\cite{ronanki2024requirements}, and scenario descriptions~\cite{zhong2023languageguided}. As LLMs have demonstrated their capability in tasks involving natural language and code generation, their application in automating code generation is promising. However, their capabilities have been examined in simple coding tasks~\cite{guo2024deepseek, chen2021codex, Liu2023CorrectCode} and not in safety-related complex applications.
Moreover, due to LLMs' known weaknesses and the safety-related nature of ADS, the generated code must follow stringent processes prescribed in ISO 26262~\cite{ISO26262} and ISO 21448~\cite{sotif} such as code reviews, verification, and validation. Software in the Loop (SiL) and simulation~\cite{fei2024SiLesmini} environments can be used to verify the code in a closed loop before it is integrated into hardware and reviewed by engineers. This serves as a preliminary validations to increase efficiency and improve code quality before other resource-demanding steps, such as code reviews.

The rise of Generative AI (GenAI) has sparked interest across many different field of knowledge. Specialized code generation with LLMs for autonomous driving and automotive applications is a less explored, yet, an increasingly popular field of study \cite{Liu2023EmpiricalStudy, ma2024lampilot, ishida2024langprop}.
Our goal is to design, implement, and evaluate an LLM-based automated code generation prototype: a pipeline aimed at accelerating automotive function development—such as ADS—in an industrial setting. Moreover, as LLMs are the main component of the designed pipeline, it is crucial to evaluate their performance with different models.
By doing so, we address the following research questions:

\begin{description}
\item[RQ1:] What are limitations of an LLM-based automatic code generation for safety-related, automotive systems?
\item[RQ2:] What are potential automatic approaches to tackle those limitations and improve the quality of the code?
\item[RQ3:] How do state-of-the-art LLM models perform under the proposed approach, when evaluated against quantitative key safety-related performance indicators?
\end{description}

The paper is structured as follows: Sec.~\ref{sec:relatedwork} summarises related work and discusses the identified research gaps addressed in this paper. Sec.~\ref{sec:methodology} presents the methodologies employed to answer the research questions. Sec.~\ref{sec:Artifact} describes the industrial use case and the key performance indicators expected from the artifact. Sec.~\ref{sec:PI_LLM_Code_Gen} presents the identified limitations of LLMs, which are considered when designing and implementing the prototype, as described in Sec.~\ref{sec:TD1}.
The evaluation results of the implemented prototype for multiple LLM models are then reported and discussed in Sec.~\ref{sec:evaluation_tests}, and final remarks and conclusion are presented in Sec.~\ref{sec:conlusion}.

\begin{figure}
  \centering
  \includegraphics[width=0.4\textwidth]{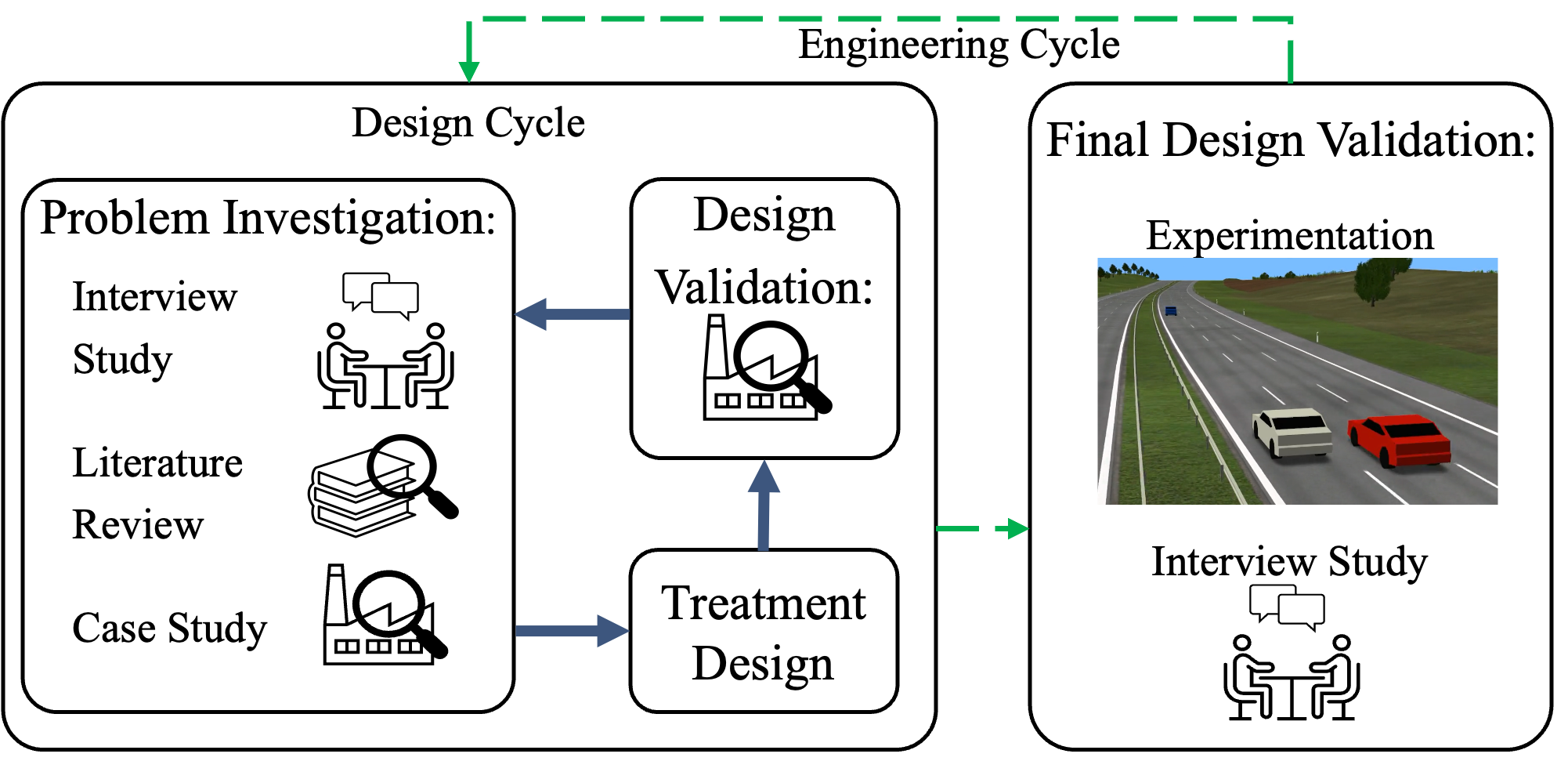}
  \caption{Presenting the design and engineering cycles in this study: The research goal is investigated in an industrial setting together with a literature review, and in the final stage, validated by an interview study. After multiple design cycles, where the designed pipeline (treatment) was improved and the generated codes, simulations, and test case reports were closely monitored. In the engineering cycle, the final treatment was validated through multiple experiments and evaluated by industrial experts in an interview study.}
  \label{fig:DesignScience}
\end{figure}

\section{Related Work}
\label{sec:relatedwork}

LLMs are used in various tasks to assist software engineers, including code comprehension and summarization~\cite{kumar2024code}, acting as chatbots for analyzing or extracting data from repositories~\cite{abedu2024llm}, and performing automation tasks at the repository level~\cite{bairi2023codeplan}.
LLMs have also been employed in automated vulnerability fixing~\cite{sagodi2024reality} and code repair~\cite{Liventsev2023} to efficiently fix software bugs without human intervention.
 \cite{Liventsev2023} explored the iterative Synthesis, Execution, and Debugging (SED) approach to leveraging LLMs in programming and overcoming the so-called \textit{near miss syndrome}.
Another approach proposes that the user break the task into sub-tasks so the model can deliver code for complex requests\cite{nijkamp2023codegen}. 

The capabilities of LLMs have also been studied in test generation, test correctness, test coverage, and test smells against existing benchmarks~\cite{siddiq2024using}. 
Benchmarks are a starting point for evaluating the performance of LLMs. However, as highlighted by~\cite{zhou2023don}, the lack of transparency regarding the training datasets of state-of-the-art LLMs raises concerns about the possibility that the benchmarks data leaked to training datasets (so-called benchmark leakage). 

\lesson{{\textbf{Benchmark leakage}} creates the risk of assessing the model's memory rather than its ability to handle unseen tasks, which is crucial for real-world applications. Therefore, as proposed in Sec.~\ref{sec:Case_Function} and reported in Sec.~\ref{sec:evaluation_tests}, new evaluation methods are needed to ensure that LLMs can meet expectations when employed for real world use-cases.}

LLMs are also employed to develop robotic~\cite{lin2023text2motion} and driving code policies~\cite{ma2024lampilot}, or to translate user intent in natural language into code~\cite{nijkamp2023codegen}.
Liu et al.\cite{Liu2023EmpiricalStudy} explored LLM-based safety-related code generation for vehicle behaviour and proposed prompt engineering methods to enhance the quality of the code. They also prescribed a manual verification and code compliance process for generated codes by LLMs.
However, manual verification comes with the cost of labor and the involvement of experts, particularly if multiple iterations are needed, as the generated code might not match the minimum expected quality.
Hence, automated verification techniques are crucial to sanity-check and bug fix in the generated code before being presented to the user. These techniques make the review process more efficient and effective, as discussed in Sec.~\ref{sec:Feedback_loop}.

\section{Methodology} \label{sec:methodology}

As shown in Fig.~\ref{fig:DesignScience}, this paper employs Design Science to design, implement and evaluate a prototype capable of automatically generating and evaluating safety-related code for automotive functions such as ADS. This pipeline, leveraging GenAI, includes mechanisms to evaluate the generated code, allowing for iterative improvements to the final function delivered to the user. The paper also details the evaluation results of this artifact. To do so, the study adhered to the guidelines by Hevner et al.~\cite{Hevner_2004}, which outline iterative phases of design performed until the objectives are met.

First, the problem is identified and the goals of each of design cycle are defined accordingly. The initial design cycle of this process, and the first step of this study, involved identifying relevant challenges in generating code using GenAI. This was done by reviewing existing academic literature and drawing from experiences within the case company, which provided the real-world context for evaluating the proposed artifact. These insights, introduced in Section \ref{sec:relatedwork} and discussed in Section \ref{sec:PI_LLM_Code_Gen}, underscore the need for the proposed artifact and address Research Question 1 (\textbf{RQ1}).

The defined objectives for each design cycle are addressed to improve the artifact, which is then tested and evaluated, with the results communicated. This process helps identify further limitations to address in subsequent cycles. Thus, in the second step, using the gathered information, the study proposed an initial version of the artifact. This first proposal was refined through iterations, such as exploring prompt engineering and other techniques, to achieve better results in terms of passing predefined tests, as described in Section \ref{sec:ScnearioandTestCases}. The various strategies explored to tackle the limitations are presented in Sections \ref{sec:TD1} and evaluated in Sec.~\ref{sec:ExperimentResults} and Sec.~\ref{sec:interviews}, addressing Research Question 2 (\textbf{RQ2}).
As the third step, the proposed safety requirements in Sec.~\ref{sec:Safety_Requirements} and test cases in Sec.~\ref{sec:ScnearioandTestCases} to evaluate and compare the generated codes by the state-of-the-art LLM models in the proposed prototype are reported and discussed in Sec.~\ref{sec:evaluation_tests}, addressing the Research Question 3
(\textbf{RQ3}).

Our study describes the final designed artifact in detail, and it presents the chain of evidence from observations in each iteration to the findings in Section \ref{sec:DE1}. In addition, we share lessons learned during the artifact development, which could guide future research on GenAI for code generation within the automotive industry.

\subsection{Literature review} \label{subsec:LiteratureReview}

The literature review was conducted using \textit{Google Scholar}, \textit{IEEE Xplore}, \textit{ACM Digital Library}, and \textit{ScienceDirect}. Relevant papers were found using the following search string, and the review was extended using forward and backward snowballing~\cite{Kitchenham2007}:

\begin{lstlisting}
    ( "active safety"
     OR "autonomous driving"
     OR "autonomous driving systems" )
AND ( "large language models"
     OR LLM 
     OR "Generative AI" )
AND "vehicle simulation"
\end{lstlisting}

\subsection{Case Study: an industry-relevant context}

The core of this research is to address an industry-relevant challenge involving complex functions, processes, and regulations. However, for novel technologies such as LLMs, the knowledge is not yet publicly documented as standards or regulations, and industrial experts are the primary source of knowledge. Thus, it is crucial to assess the validity of the proposed treatments in an industry setting by employing case studies and interviewing experts in the industry to collect their feedback. The case study is conducted in a company involved in ADS development, and the results are presented and discussed both in the case company and their supplier software company involved in ADS development.
The developed prototype is designed to be suitable for use in real industrial settings, such as inputs to the pipeline, simulation model, and acceptance criteria. This setup ensures that the designed prototype is a practical tool that aligns with industry practices, needs and requirements.

\subsection{Experiment setup: esmini}\label{subsec:ExperimentSetupesmini}

The prototype was implemented by integrating Environment Simulator Minimalistic (esmini)~\cite{esmini_user_guide}, as an abstract world model in which the function could operate and be visualised, along with a report generator. esmini is considered a suitable environment because our prototype aims to generate code for the function abstraction level in the concept phase, which aligns with its capabilities~\cite{fei2024SiLesmini}.
The generated software in each iteration is tested in the virtual environment, and reports are generated automatically. esmini provides a realistic testing environment by allowing the generated code to control the ego vehicle in multiple critical scenarios.
As this simulation environment is used in the verification and validation of both ADS and advanced driver assistance systems (ADAS) functions in the case company, it not only becomes part of the pipeline (providing feedback to the LLM) but also serves as an intuitive evaluation tool for the implemented prototype in this study.
Thus, the quantitative and qualitative pass or fail generated reports are used as Key Performance Indicators (KPIs) for evaluation of proposed treatments in each engineering cycle.

\subsection{Triangulation with Interview Study}\label{subsec:MethodInterviewStudy}
As the final evaluation stage, 12 experts from two OEMs involved in ADS development are interviewed, and their comments and insights are reported in Sec.~\ref{sec:interviews}.

\emph{Interview Method:}
The interviews were conducted through Microsoft Teams and ranged from 30 to 60 minutes. 
To ensure consistency in data collection and to guide the researcher, a detailed interview protocol was designed after the pilot study\footnote{Interview study protocol: \url{https://doi.org/10.5281/zenodo.14783284}}.
During the interview, a Microsoft Forms containing all the questions was filled out and then sent back to the experts after the interview for validation.
The demo of the prototype was pre-recorded and played for the participants. The demo included the generated code, corresponding visualised simulations, and test reports for multiple iterations of the prototype.
After the demo, the experts could ask questions, and the interviewers could provide clarifications.

\emph{Questions:}
The experts expressed their opinions through open-ended and multiple-choice questions, followed by optional comments to encourage and allow for open-ended discussions. The researcher could also ask additional questions based on the experts' answers to further explore their insights.
The experts were initially asked to express their experience in using LLMs for code generation. Subsequently, the collected challenges were presented to them as statements, and they could select between strongly agree, agree, disagree, strongly disagree, and neutral. Neutral could be selected if they had no opinion or preferred not to answer. After each challenge, there was an option to add further comments for clarification or to share their experiences.
In the final stage, the usefulness of the tool was explored, including its advantages and disadvantages, weaknesses and strengths, potential ideas for improvement, and other use cases of the implemented prototype.

\emph{Participants:}
To be eligible, the participants had to have at least 5 years of experience in ADS development. 11 experts were selected from 2 OEMs with diverse roles to validate and explore both the problem relevance and the prototype designed for it. The group included one manager, three product owners, two verification and validation engineers, one data scientist, one system architect, one programmer and three system safety experts.
The experts had on average 12 years of experience, and were located in three different countries. An ID (Px) is assigned to each expert, and they are referred to in the text using their ID.
The ethical guidelines for software engineering interview studies, such as obtaining consent, anonymization, and confidentiality, as proposed by  Strandberg~\cite{swethic} were followed.

\section{Artifact}\label{sec:Artifact}

\subsection{Industrial Use Case and Requirements on Prototype} 
The needs and requirements from the case company were collected during the initial engineering cycles through continuous discussions, considering both the OEM's needs and the capabilities and limitations of LLM technology. Below is the final use case description of the prototype, along with the requirements.

\emph{Use Case:}
The main use case of the prototype is 
in the concept phase and at the function abstraction level for verifying and validating the function description. It also helps with the development of a formal notation (i.e., Python code) of the function description in natural text to better explain the logic of the function, and finally, visualizing the logic. Both the logic and visualization help to capture safety-related weaknesses in function requirements, such as contradictions between functional requirements or ambiguities.
The pipeline shall provide the code, simulation logs, visualization, and generated test report to the user. After the user reviews these materials, the selected code and other relevant artifacts are attached to the function description as supplementary materials for the next steps in the development process.

\emph{Inputs:} The only allowed inputs to the prototype are the function description in natural language, scenario descriptions in ASAM OpenSCENARIO~\cite{asam-openscenario} and OpenDRIVE~\cite{asam-opendrive} format, and test cases. The prototype shall create the first version of the code, with no need for legacy code. Although the user can provide hints in the function description to guide the LLM, the prototype shall be evaluated without legacy human-generated code.

\emph{Efficiency and Efficacy:}
The pipeline needs to continue the iterations until all test cases are passed or the maximum allowed number of iterations is reached. The report shall summarise the passed and failed test cases so the user can rapidly assess the maturity level of the code. This enables the user to decide more quickly whether the code is ready for final review and bug fixing, and what needs to be improved.

\emph{Automation:} The prototype shall automate all tasks without human engineer intervention. This includes calling the APIs, and managing communication between the LLM, simulation, and report generator.

\emph{Programming language:} The simple syntax, semantics, and high-level constructs of Python make it an ideal option for testing function abstraction levels. Additionally, Python has an average Copilot output quality (42\%) (for comparison: Java~(57\%), JavaScript~(27\%), and C~(39\%)~\cite{Nguyen2022EmpiricalEvalCopilot}), it is a suitable candidate for representing the generalizability of the pipeline to other programming languages. This ensures that the findings are more broadly applicable. 

\subsection{Case Function}\label{sec:Case_Function}
Selecting the appropriate function for the experiment is crucial, as factors such as algorithm complexity and prior knowledge of LLMs are important considerations for improving the validity of the experiment.
As explained in the motivation, the main potential use case of the prototype is for function expansion and rapid reaction to identified unknown scenarios. Therefore, the function selected for evaluating the effectiveness of the prototype is the delta between version one and version two of ADS, as proposed by ISO/TR~4804~\cite{isotr4804}.
``Unsupervised Traffic Jam Chauffeur System'' (Version one) is only operational when there is a lead vehicle and shall stay in the same lane~\cite{isotr4804}. If a lane change is needed, it must be performed by the driver. The vehicle can avoid collisions only by braking and is not allowed to perform evasive manoeuvres.
\emph{Motorway Chauffeur System (Version two)} is operational with or without a leading vehicle and is allowed to change lanes without human supervision~\cite{isotr4804}.

\emph{Unsupervised Collision Avoidance by Evasive Manoeuvre (CAEM)} is a potential add-on function to version one of ADS (i.e., lane change not allowed), moving towards version two (i.e., lane change is allowed). It is responsible for performing unsupervised lane changes when braking cannot avoid a collision with vehicle in front due to a braking system failure, or insufficient braking distance.

In this study, other functions such as Automated Emergency Braking (AEB) and Adaptive Cruise Control (ACC) are also used as case functions, although they are not used as the main criteria to evaluate the performance of the prototype due to the lower complexity of these two functions. Moreover, as AEB and ACC are mature, there is plenty of data (e.g., code, and documentation) in public databases, which increases the probability of the LLM being trained on them, posing a threat to validity for evaluation~\cite{nouri2024RELLM}.
On the contrary, CAEM is not only more complex but also quite immature with respect to existing publicly available data.

\begin{figure*}
  \includegraphics[width=\textwidth]{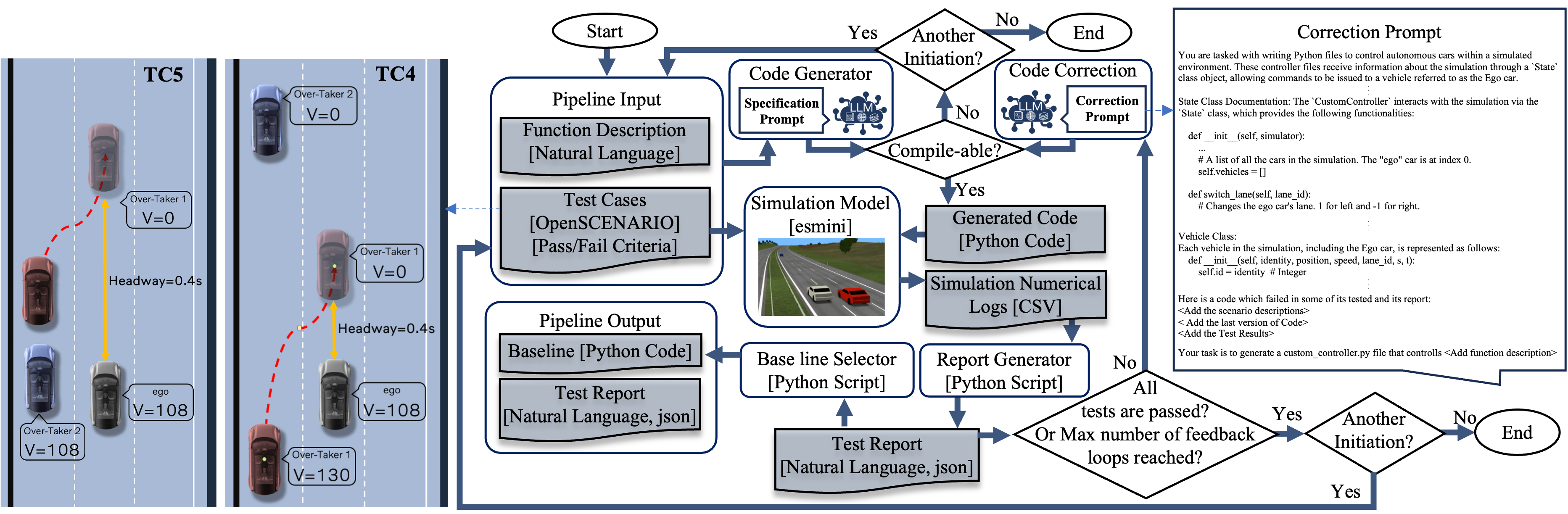}
  \caption{The implementation for iterative automated LLM-based code generation using a simulation model for safety evaluation and for improving the generated code: The pipeline receives the function description and test cases (e.g., TC4 and TC5 depicted on the left side), and safety acceptance criteria. Initially, the pipeline uses a Specification Prompt to generate the first version of the code. This code is then sent to the simulation model, and a test report is generated in natural language. This report is used in the Correction Prompt (template on the right side) to generate subsequent versions of the code based on the initial version. The loop (Simulation-LLM conversation) continues until the code passes all test cases or reaches the maximum iteration limit. The user can also initiate a fresh start of the pipeline, generating a new controller without considering feedback from previous iterations. Each generated version of the controller is compared to the selected baseline using the generated test report.}
  \label{fig:LLMPipeline}
\end{figure*}

\subsection{Safety Requirements}
\label{sec:Safety_Requirements}
Given that the hazards associated with the function can lead to safety-related losses, it is essential to specify safety requirements (SRs) that must be met to ensure its safe operation. The safety requirements are specified as following:

\begin{description}
\item[SR1:]
 Avoid collisions with dynamic or static objects on the road.
\item[SR2:] Avoid exiting drive-able area.
\item[SR3:] Avoid unintended lane change if there is no imminent collision with the vehicle in front.
\end{description}

\subsection{Test Cases}
\label{sec:ScnearioandTestCases}
The safety of the generated code is evaluated based on the number of successfully passed Tests Cases (TC) each covering one or multiple safety requirements. 

\textit{Cut-in \& decelerate at 120, 80, \& 40 kph (TC1, TC2, and TC3):}
These three scenarios occurs on a multi-lane highway, with the ego vehicle driving with constant speeds of 120 kph (TC1), 80 kph (TC2), and 40 kph (TC3) in the second lane from the left, while the Over-Taker vehicle starts in the adjacent lane (first lane), matching and slightly exceeding the ego vehicle’s speed. The Over-Taker performs a cut-in manoeuvre and immediately decelerates, while the ego vehicle maintains the speed. According to UNECE Regulation No. 157~\cite{ALKS}, time gap from having perceived the braking by the front vehicle to starting reaction (e.g., braking) by the ego vehicle, is 0.75 seconds for a skilled human driver~\cite{ALKS}.
On the other hand, for those three scenarios, the time gap between the Over-Taker and the ego vehicle (i.e., headway time) at the moment of deceleration is only 0.4 seconds, which is significantly shorter than the 0.75 seconds in terms of reaction time, making the scenario more challenging.

\textit{Cut-in \& decelerate with Lane Blockage (TC4 and TC5):}
As depicted in Fig.~\ref{fig:LLMPipeline}, the Over-Taker in these two scenarios behaves the same as in previous scenarios, with the ego vehicle maintaining a speed of 108 kph. However, in these scenarios, an additional vehicle serves the role of blocking the left lane. In TC4, the static vehicle is placed further ahead in the left lane, which is the proposed lane for the ego vehicle to perform an evasive manoeuvre. This could lead to a collision if the ego vehicle does not respond appropriately. In TC5, the blocker adjusts its speed to match the ego vehicle when it reaches the left of the ego vehicle, blocking the lane.

\textit{Unintended lane change (TC6 and TC7):}
In the last two scenarios, the absence of unintended lane change behaviour (i.e., false positive) is tested.
In TC6, there are no other vehicle on the road to check if the controller performs any unintended lane changes. In TC7, the ego vehicle is in the first lane from the left while another vehicle passes in the opposite direction.

To more accurately assess the safety of the generated controller for CAEM, it is assumed that the vehicle maintains a constant longitudinal speed (i.e., no deceleration) during the collision avoidance, which increases the difficulty of the scenario compared to real-world conditions. Hence, in TC1 to TC5 the function shall perform lane change to satisfy \textit{SR1}. Additionally, in scenario TC4 and TC5, the function need to avoid collision with the second agent as performing the lane change. 
In all scenarios the controller shall stay on the road to satisfy \textit{SR2} which is challenging especially while performing the evasive manoeuvre.
TC6 and TC7 are designed to test if the generated controller violate \textit{SR3}.  

\section{LLMs limitations in Code Generation} \label{sec:PI_LLM_Code_Gen}

A number of limitations of LLM-based code generation tools were extracted from the literature and interviewees' insights.
Said limitations are reported in this section, clustered based on the proposals in the literature and their importance in safety-related domains. 

\textit{It is difficult to evaluate generated code} 
Code review and test are required steps in safety-related software before its release~\cite{Nouri2022SEAA}. However, Vaithilingam et al.~concluded that the ``participants often failed to understand and assess the correctness of the generated code''~\cite{Vaithilingam2022ExpectationVSExperience}. This might be because LLMs often generate hallucinations and use uncommon APIs~\cite{aleti2023software, liang2023largescale}. Moreover, lower readability and comprehensiveness of LLM code have also been reported~\cite{Jimenez2024swebench}, which can lead to failing to find bugs.
Surprisingly, interviewees P3, P5, and P11, who are experienced in reviewing LLM-generated codes, disagreed with this limitation, and stated that LLM-generated code is easy to understand, even more than human-generated code, according to P5. 
Interviewees P1, P10, and P11 added that the reviewers' knowledge and experience is an influential factor in failing to detect a software fault~\cite{xu2021inide},

\textit{Required review and bug fixing effort} 
It has been reported in the literature that reviewing, bug-fixing, and modifying the generated code requires too much time and effort~\cite{Vaithilingam2022ExpectationVSExperience, liang2023largescale}, which was confirmed by three of the interviewees: P6, P7 and P8. However, they judged it similar or even easier to review than human-written code. 
The issue might however lie in LLMs' tendency to generate large amounts of code quickly without guaranteeing the minimum functional or non-functional acceptance criteria, which can lead to time and resource-intensive review efforts. These efforts could be reduced by improving the quality of the code using a closed-loop feedback mechanism to the LLM and iterating on it. To mitigate this, novel testing methods, covering both functional and non-functional requirements of the code, could be used to perform preliminary validations before the engineer conducts a code review, reducing the number of iterative reviews.

\textit{Prompt Engineering difficulties} 
Interviewee P1 stated that communicating with an LLM is an engineering skill, that prompts are a programming language, and need to be more specific.
For instance, ambiguous prompts can lead to errors in the generated code~\cite{poldrack2023aiassisted}.
Code-generation prompts might therefore require detailed explanations, which might exceed token limitations, or breaking the tasks into sub-tasks or sub-functions. However, as P4 and P6 pointed out, missing context in the prompts are potential reasons for lower performance, as is using outdated models. This might lead to the generated code not satisfying all requirements in the prompts~\cite{liang2023largescale}.

\textit{Unnecessary complexity of generated code} 
It has been reported that LLM-generated code is unnecessarily more complex~\cite{Nguyen2022EmpiricalEvalCopilot} and less efficient~\cite{Jimenez2024swebench}. Surprisingly, none of the interviewees agreed, and countered that the complexity depends on the task, as does for human-written code.

\textit{Incorrect understanding of domain-specific terms} 
LLMs might not understand the intended definition of some domain-specific terms~\cite{Liu2023EmpiricalStudy, nouri2024RELLM}, which can lead to wrong interpretation of the prompt. All interviewees agreed to this limitation, and shared their experiences related to understanding specific terms from ISO 21448, and specific communication protocols. Training the model using domain-specific data was among the proposed solutions.

\textit{Incorrect, inaccurate, insecure, or unsafe code} 
The so called ``near miss syndrome'' or ``last mile problem'' refers to generations that are close to correct but not entirely operational~\cite{Liventsev2023}, recommend compromised packages, or use insecure function calls~\cite{chen2021codex}.
Moreover, it is hard to predict the LLM-generated code~\cite{liang2023largescale,aleti2023software}, specially for some programming languages~\cite{Nguyen2022EmpiricalEvalCopilot} and for some tasks. 
In this regard, interviewee P6 shed light on GPT4 struggling to generate code for new tasks, while recognizing its usefulness for tasks with existing code. 
All experts, however, highlighted similar challenges in human-generated code, and suggested using similar testing and code review mechanisms as those used today.

\textit{Intellectual Property (IP) infringement} 
LLMs' training may include copyrighted material, which might reappear in the generated code (even if partially) and might not be identified by code reviewers~\cite{Li2022AlphaCode,chen2021codex}.
For this reason, a fifth of participants in a survey stated not using AI programming assistance to avoid potential IP infringement~\cite{liang2023largescale}. 
According to P10, this is a real concern that poses legal risks for companies. 
On the other hand, providing requirements and code to LLMs might expose users' or their company's IP to third parties.
Confidentiality was however not a concern for most interviewees, as some providers guarantee the confidentiality of conversations and to not use user data for model training.

\textit{Not suitable for complex software} 
LLMs might not be able to handle or generate correct code for complex algorithms~\cite{Li2022AlphaCode}.
According to P3, GPT-4 Turbo is quite powerful for generating short snippets of code as long as the logic is ``shallow,'' even though the maturity of the model could be an influential factor.
Decomposing complex tasks is not only seen as an approach for improving LLM-generated code, but  also for effective and efficient reviewing processes.

\section{Treatment Design}
\label{sec:TD1}
To satisfy the requested use case (Sec.~\ref{sec:Artifact}) using GPT-4, considering its limitations as identified in Sec.~\ref{sec:PI_LLM_Code_Gen}, a closed-loop iterative LLM-based pipeline is proposed and a prototype is implemented. As presented in Fig.~\ref{fig:LLMPipeline}, this pipeline leverages code execution through simulation and creates test reports in natural text as feedback to the LLM to improve the generated code.
The concept builds on the idea that the initially generated code, which might not be fully functional or may violate some safety requirements such as avoiding collisions, can be improved through additional iterations by providing feedback. The feedback highlights details regarding malfunctions~\cite{ISO26262} or functional inefficiencies~\cite{sotif} of the generated code, which are translated from simulation logs into a natural language report and provided to the LLM for iterative bug fixing and code improvement.
The concept of enhancing the code generation model in the artifact through targeted evaluation and refinement is associated with previously examined ideas in the literature such as \textit{iterative improvement}, \textit{learning from errors}, and \textit{multi-turn} generation~\cite{nijkamp2023codegen, Huang2023FineTuneAPR, xia2022training, Liventsev2023, shinn2023reflexion}. 
The designed prototype receives the function description of the desired ADAS or ADS function in natural language, along with traffic scenario files and test cases to evaluate the generated code. It then provides the generated code along with a test report in natural language.

\subsection{LLM and prompt engineering:}

To address research questions RQ1 and RQ2, the initial problem investigation and validation involving standard experiments are done to assess the capability of GPT-4, like the one that is done in hiring process. As a result, the capabilities and limitations of the GPT-4 and weaknesses of engineered prompts are identified. Then the prompts are improved to tackle identified issues.
Liu et al.~\cite{Liu2023EmpiricalStudy} reported on a set of prompt engineering techniques for code generation, such as constraints on the code, detailed step descriptions, and inclusion of examples in the prompts, in addition to the primary functional description of the code. Guidelines from standards and regulations can be included as constraints in the prompts.
Disambiguation of the prompts can be done by integrating the local context of the users, which can lead to improvements in the output~\cite{xu2021inide}.
Initiation of the prompts using context is also examined in~\cite{nouri2024RELLM}.
Moreover, domain-specific terms and requirements need to be transformed to general requirements~\cite{Liu2023EmpiricalStudy}, or described as a separate context at the start of the prompt~\cite{nouri2024RELLM}.
During the initial steps, manual test reports are provided as feedback to GPT-4 to better understand the effective prompt strategy for providing feedback and observing the evolution of software. Identifying the required information in the reports provided to GPT-4 is also another outcome of this step.

There are two categories of prompts: the initial prompt to generate the controller for the first time without any feedback (Specification Prompt); and the second one, for generating subsequent versions based on the initial version (Correction Prompt).
The Specification Prompt consists of \textit{Context} and \textit{Task Description}.

\textit{Context} defines the required programming language and the generic coding instructions necessary for integrating the generated controller with other software components in the system. In this study, the code interacts with a simulation model, which includes instructions on how to read the state of the ego vehicle and other agents within the simulation. This section is a generic description that remains the same for any new function.

\textit{Task Description} provides a description of the function, commonly referred to as the item definition in the automotive domain, which specifies the abstract behaviour required from the function. The task description may also include known legal or safety requirements.
For instance, in this case, the LLM is instructed to prioritise lane changes to the left.
Additionally, it can incorporate lessons learned and suggestions. For example, Time to Collision is defined in natural language as a potential metric: "A measure for imminent collision is the time to collision (TTC), defined as\dots". Known logics can also be represented as pseudo-code; however, this was not used in the study to maintain the validity of the experiment.

As presented in Fig.~\ref{fig:LLMPipeline}, the Correction Prompt consists of \textit{Context}, \textit{Scenario Description}, \textit{Last Version of Code}, \textit{Test Results}, and \textit{Task Description}.
\textit{last version of Code} is the previous version of the generated controller code, which is provided for creating the updated controller.
\textit{Scenario Description} is the natural language description of each scenario. \textit{Test Results} includes descriptions of both passed and failed test cases, which are generated automatically by a Python script based on the logged data from simulation.
For instance, in one case that the ego vehicle collides with another vehicle, the result is stated as: ``Ego was involved in a collision at time: 13.3 seconds with a speed of 33.33 m/s, colliding with: OverTaker.''

\begin{figure}
    \centering
    \includegraphics[width=0.8\linewidth]{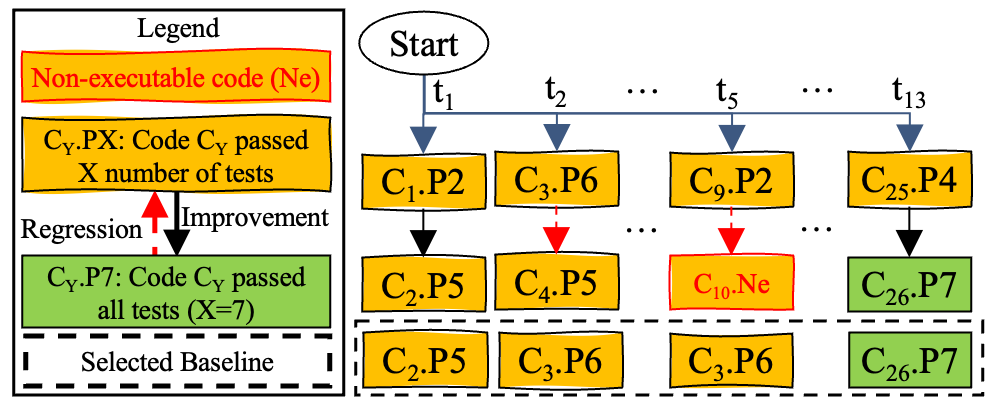}
    \caption{Illustration of the baseline selection strategy across generated code versions over multiple iterations of the proposed feedback mechanism at step $t_n$. The code version $C_{2n-1}$ represent the initial version in each cycle, generated without any feedback. Through refinement via the feedback mechanism, an enhanced version $C_{2n}$ is generated and evaluated automatically. The newly generated version is then compared with the baseline on the number of successful test cases, which serves as a measure of the code’s robustness and safety in handling safety-related scenarios. 
    The pipeline can continue this process both horizontally (i.e., generating new code) and vertically (i.e., refining existing code based on generated test report) until the code passes all tests (e.g., $C_{26}$), referred to as the gold baseline, or reaches a maturity level (e.g., $C_{3}$) that can be further refined by the engineer. There may also be instances where the generated code is non-executable (e.g., $C_{10}Ne$) for one or more test cases in the simulation due to runtime or syntax errors.}
    \label{fig:ChainOfThoughts}
\end{figure}

\subsection{Feedback loop}
\label{sec:Feedback_loop}
The generated code is extracted from the received message from LLM and is tested against various safety-related traffic scenarios, as discussed in \ref{sec:ScnearioandTestCases}, to identify safety-related hazards caused by the controller and report them back to the LLM. 
For each traffic scenario, multiple acceptance criteria, such as collision detection, were evaluated.
As a result, the esmini simulator provides tabular log files detailing the positions and velocities of each vehicle at each time-step for every test case. Since the LLM is not able to reliably interpret these log files, a report generator was developed to translate the numerical logs into natural language and report them, summarising the pass or fail state of all acceptance criteria for each test case, forming a Simulation-LLM conversation.

In each iteration, if the code passes all tests or the specified iteration limit is reached, the code generation process stops, and the prototype exits the loop. 
There is also the option to start fresh and explore newly generated codes, which has been found to be effective when combined with the feedback loop.
Each generated code is compared to the current baseline based on their respective test reports. The best-performing code is selected in each loop as the new baseline and delivered along with the corresponding report.

\section{Evaluation and Validation of Pipeline} \label{sec:evaluation_tests}

Initially, in Sec.~\ref{sec:ExperimentResults}, the performance of the pipeline is evaluated for the main configuration (Model: GPT-4, Function: CAEM). Subsequently, in Sec.~\ref{sec:evaluation_Models} the performance of other configurations of the prototype is evaluated and compared. These configurations involve various open-source LLM models~\footnote{deepseek-coder:33b (ID: acec7c0b0fd9); deepseek-r1:32b (ID: 38056bbcbb2d);  codegemma:latest (ID: 0c96700aaada); mistral:7b (ID: f974a74358d6); codellama:34b (ID: 685be00e1532)} and GPT-4 for two functions (ACC and CAEM).
Lastly, the effectiveness and efficiency of the prototype were evaluated through an interview study by demoing the tool to experts involved in ADS development, as explained in Sec.~\ref{subsec:MethodInterviewStudy}. Moreover, the assumed challenges of the employed technology, the use cases of the artifact, and the proposed treatments were validated. The diversity of the experts, in terms of roles and experience, helped to better explore potential limitations, weaknesses, use cases, or concerns related to both the LLM and the pipeline.

\subsection{Evaluation of Pipeline by Experiment}
\label{sec:ExperimentResults}

\begin{figure}
    \centering
    \includegraphics[width=1\linewidth]{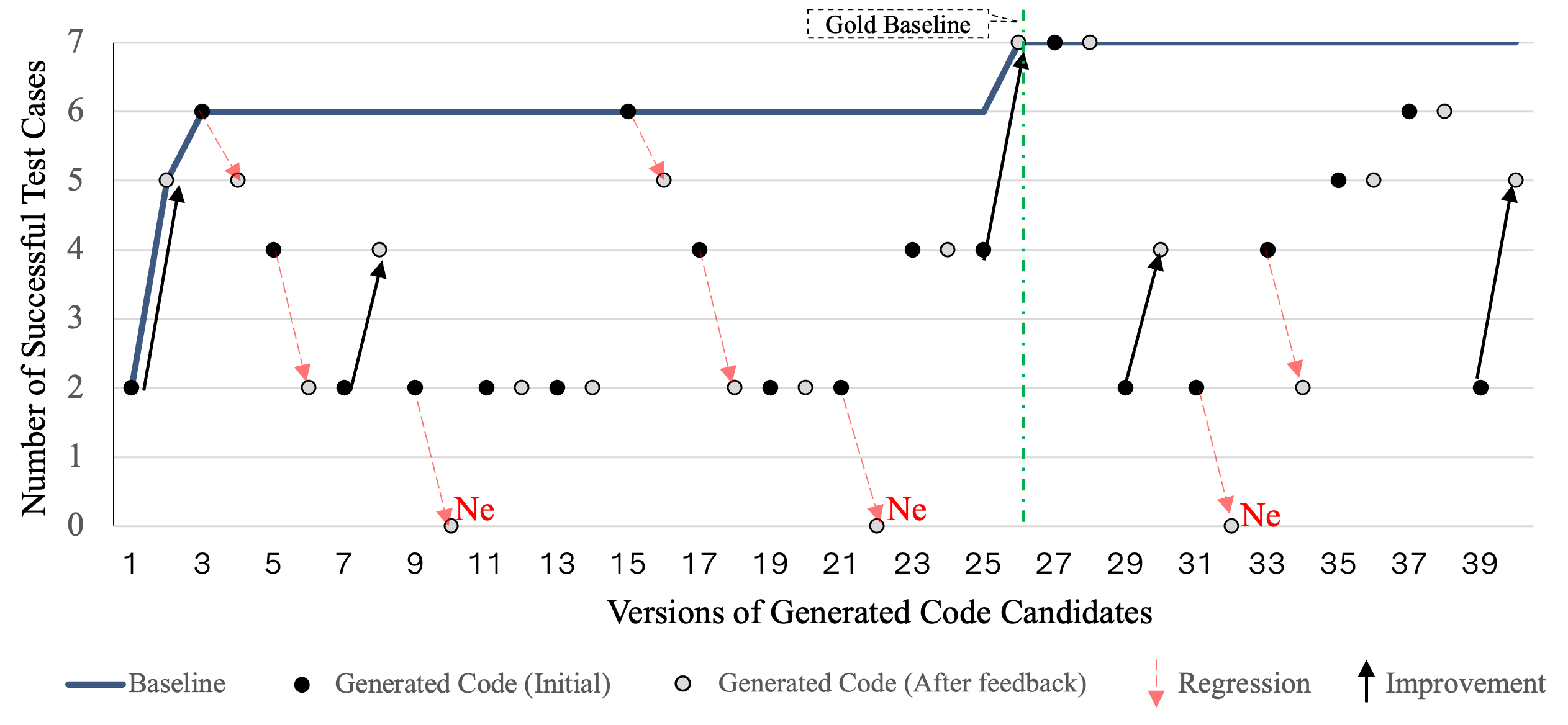}
    \caption{Visualization of the progression of generated codes and their selected baseline at each step. Black arrows show performance improvements through simulation-based feedback to LLM, while red dashed arrows highlight regressions between initial versions and subsequent refinements. The gold baseline (e.g., ($C_{26}$)), indicating the version that passed all test cases successfully.}
    \label{fig:Results_all}
\end{figure}

As presented in Fig.~\ref{fig:LLMPipeline} and illustrated in Fig.~\ref{fig:ChainOfThoughts}, the pipeline generates one code per initiation.If needed, the code is sent for correction, leading to an enhanced version. As a result, two versions of the code are produced per initiation. In this experiment, the pipeline was initiated 20 times, resulting in 40 versions of the code for the CAEM function using GPT-4.
These versions were tested across multiple traffic scenarios simulated in esmini, as discussed in Sec.~\ref{subsec:ExperimentSetupesmini} and~\ref{sec:ScnearioandTestCases}. 
Each version of the code is then analyzed based on the number of successful test cases that satisfy the safety acceptance criteria, and the results are presented in Fig.~\ref{fig:Results_all}. 

As shown in Fig.~\ref{fig:Results_all}, the baseline achieved 6 successful test cases by the third version, while the golden baseline, which satisfies all test cases, was reached in version 26. In the 13th fresh start of the pipeline ($t_{13}$), the initial code version ($C_{25}$) passed 4 test cases. Through the feedback mechanism the next version ($C_{26}$) reached the mature stage and successfully passed all test cases\footnote{Appendix providing the sample generated codes and reports for ($t_{13}$) and the simulation visualisations, \url{https://doi.org/10.5281/zenodo.14783374}} \footnote{Video demonstrations of scenarios, generated code (initial and corrected loops), and test reports are available at \url{https://youtu.be/cfmS8BP1KPI}}.
There are three instances in which the attempted correction could not be executed in the simulation (e.g., $C_{10}$), and three additional instances that result in run-time errors for some scenarios (e.g., $C_{18}$).
These six cases lead to regression compared to the initial version and are flagged by the pipeline, as illustrated $C_{10}$ in Fig.~\ref{fig:ChainOfThoughts}. After accounting for corrections that resulted in a second fully executable version, there were 14 initial generations and 14 enhanced versions, the effectiveness of the feedback loop and correction prompt can be evaluated. The results show an average improvement of passed tests cases (P) of 9.2\% of all 14 corrections and 37\% considering only the improvements (5 corrections).



\lesson{{\textcolor{BlueViolet}{\textbf{Insight:}}} In the correction attempt, regressions might occur; however, the pipeline detects and removes them.}

\subsection{Evaluation and Comparison of LLM Models} \label{sec:evaluation_Models}

\begin{figure}
    \centering
    \includegraphics[width=1\linewidth]{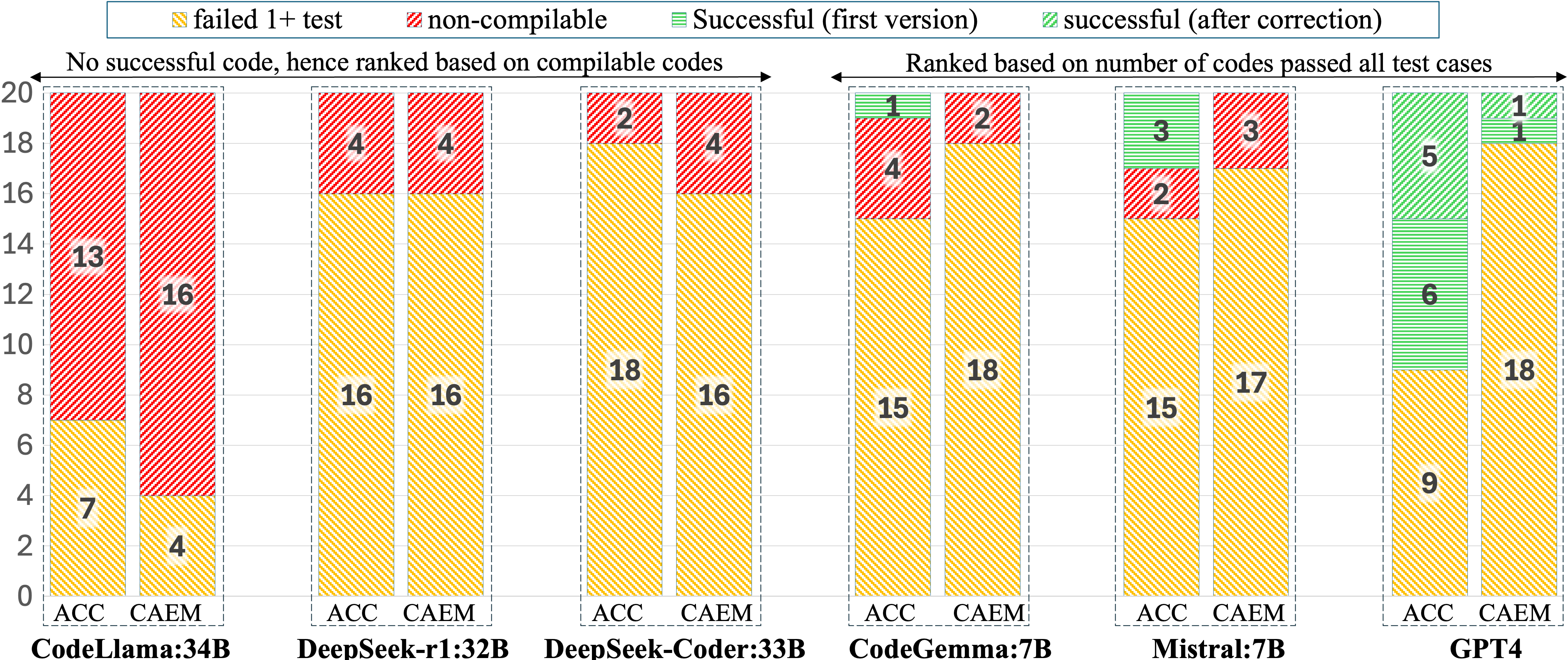}
    \caption{Performance comparison of different LLM models in the designed pipeline for ACC and CAEM. Each model generate 20 codes for each function. Codes are checked for syntax errors (red dotted bars) and then evaluated on all test cases. If the code passes all test cases, the process is completed (green bars with horizontal lines). If the compiled code fails one or more test cases (orange bars with diagonal lines), it is sent to correction attempt and tested if lead to a correct code (green bars with diagonal lines). 
    The models are ordered from left (lowest performance) to right (highest performance) based on two criteria: (1) the number of codes that passed all test cases and (2) if none of the codes are successful, then based on the number of compilable codes.}
    \label{fig:Results_all_models}
\end{figure}

As it is shown in Fig.~\ref{fig:Results_all_models}, CodeGemma:7b and Mistral:7b managed to generate successful codes for ACC (1 and 3 instances, respectively), while Codellama:34b, DeepSeek-r1:32b, and DeepSeek-Coder:33b failed.
None of these five open source models successfully generated fully functional and successful code for CAEM.
CodeLlama had the highest number of non-compilable cases, with a total of 26, primarily due to syntax errors or instances where no code was generated. In comparison, CodeGemma had 24 non-compilable cases, and Mistral had 11. Although compilable code does not guarantee successful execution (e.g., runtime errors), it can still serve as the second criterion for comparing model performance if there is no successful code. This is especially true since producing compilable code, even with incorrect behavior, is generally the minimum expectation from code generator.

\lesson{{\textcolor{BlueViolet}{\textbf{Insight:}}} Unlike Codellama:34b, DeepSeek-r1:32b, DeepSeek-Coder:33b, CodeGemma:7b, and Mistral:7b, all codes generated by GPT-4 compiled. GPT-4 demonstrated the highest success rates for both ACC (6 for the initial version and 5 after correction) and CAEM (1 for the initial version and 1 after correction).}


\subsection{Evaluation by Experts} 
\label{sec:interviews}

Seven experts in the interview study voted for ``very useful'' and the rest voted for ``useful,'' citing several reasons. The tool reduces the effort and workload of developers, saving time for new feature development (mentioned by 8 experts) and serving as a prototyping tool in concept phase (mentioned by 3 experts). It automates the entire software development process and making it faster and less resource-demanding than human coders, by adapting the code based on simulation feedback and failed tests (mentioned by 3 experts). Additionally, it helps to code complex ADS rules and requirements (mentioned by 1 expert), and employ creative strategies to enhance development efficiency (mentioned by 1 experts). Finally, it aids in rapid visualization (mentioned by 1 experts) and improvement of function description, concept or algorithms in the initial development phase (mentioned by 4 experts).

Despite its strengths, experts emphasised the necessity of having human in the loop for reviewing and testing production-related code. For instance, one expert mentioned that, since the LLM does not have the bigger picture in which the function is used, it might deviate from the intended goal, making human supervision crucial.
Another expert highlighted the importance of suitable traffic scenarios and test cases as critical components of the feedback to LLM. It was also proposed to update the input requirements (i.e., function description) in each correction loop to address the malfunctions or functional insufficiencies reported by the simulation.

\subsection{Discussion}
\label{sec:DE1}

In both configurations of GPT-4, the success rate (i.e., the number of fully successful codes divided by the total generated codes) of the correction loop is slightly higher than that of the first version. This underscores the advantage of improving failed codes rather than disregarding them and generating new ones. For example, the success rate for ACC increased from 30\% (6 out of 20) in the first version to 35\% (5 out of 14) in the correction loop. Moreover, as also reported in Sec.~\ref{sec:ExperimentResults}, the number of satisfied acceptance criteria increased on average with each correction using feedback generated from the simulation.
Another advantage of the correction loop is that, as the generated codes employ diverse strategies, it is beneficial not to disregard a strategy if it fails the first time. Instead, providing the strategy to the LLM for improvement might lead to a more diverse set of approaches, potentially resulting in safer or more efficient solutions.
Each full execution of the pipeline took less than 3 minutes from the initiation to delivering both the initial and enhanced versions, including simulations and report generation. This means that 20 full executions of the pipeline required less than an hour, which, in our experiment, led to the delivery of two fully successful versions for CAEM and 11 for ACC.

As shown in Fig.~\ref{fig:Results_all_models}, the correction loop did not improve the number of successful codes for CodeLlama, DeepSeek-r1, DeepSeek-Coder, CodeGemma, and Mistral. However, for GPT-4, the correction process led to a higher number of successful codes. 
This discrepancy likely caused by token limitations in all models except GPT-4, as correction prompts require additional tokens to accommodate the code and report.
To partially mitigate this issue, the prompts were designed to prioritise the most important parts of the task at the end of prompts.

\lesson{{\textcolor{BlueViolet}{\textbf{Insight:}}} As shown in Fig.~\ref{fig:Results_all_models}, Mistral:7B and CodeGemma:7B outperform CodeLlama:34B and DeepSeek-r1:32B, demonstrating that more parameters in an LLM do not necessarily lead to better results.}

As shown in Fig.~\ref{fig:Results_all_models}, none of the open-source models generated any fully functional and successful code for CAEM, likely due to the greater complexity of CAEM compared to ACC or the potential leakage of publicly available code for ACC, as discussed in Sec.~\ref{sec:Case_Function}.
This trend is also observed for GPT-4, as the total number of successful codes for CAEM is 2, while for ACC it is 11.

\lesson{{\textcolor{BlueViolet}{\textbf{Insight:}}} The failure of open-source models to generate functional code for CAEM indicates that these models are not yet ready for code generation in new tasks like CAEM.}


According to Liu et al. \cite{Liu2023EmpiricalStudy} there is a need for a new V model for safety-related software development using LLMs. In their proposal system abstraction levels are removed, including system design and system testing.
However, their proposal has significant shortcomings, which pose several challenges. Firstly, the existence of abstraction levels is essential for decreasing the complexity of the system, especially in systems such as ADS. As LLMs have token limitations, like human engineers, they also have limitations on complexity. Moreover, the modularization of safety-related software is needed to improve code review and is required for software allocation to different partitions. Thus, the modification of abstraction levels and processes in safety-related systems might be irrelevant to the tool used in generating the code. For instance, Nijkamp et al. \cite{nijkamp2023codegen} reported improved accuracy of the code by decomposing the main problem into subproblems, asking for a subprogram at each step, and then integrating the results.

However, the impact of using LLMs should be analysed to identify the required modifications in the activities involved in the development of safety-related functions.
For instance, requirement engineering might require new insights, as requirements are also used as input to the LLM. It is also crucial to closely monitor these tools to identify their shortcomings. This is a prerequisite for the software tool qualification process and certification of the tool for use in production-related software. Moreover, other legal concerns such as IP infringement and copyright remain possible limitations in the applicability of code generation with LLMs in the industry.

\subsection{Threats to Validity}
\label{sec:Threats}

\emph{Construction Validity:} 
The final prototype is evaluated through experimentation using a simulation model. Inaccuracy or oversimplification of simulation models relative to the real environment are critical threats to construct validity. However, since esmini is used in similar industrial setups for the same purpose, this risk is mitigated. Additionally, the purpose of the pipeline is to improve code quality by providing feedback from the model to the LLM. An interview study is employed, serving as a triangulation mechanism to validate the effectiveness and efficiency of the designed and developed pipeline, including the simulation model.

\emph{Internal Validity:} 
LLMs' stochastic behaviour in code generation might lead to variability in the quality and functionality of the generated code. This can result in inconsistencies across different iterations of the same prompt, potentially affecting the reliability of evaluation results. To mitigate this the pipeline is initiated 20 times to ensure robustness and consistency in the results.
Additionally, the chosen case function for evaluating the prototype is a novel function, minimizing the risk of data leakage from the LLM's training data. This also improves the validity of the evaluation by focusing on the model's innovation rather than its memory.
Selecting experts from diverse backgrounds and roles, conducting a pilot study, including different geographical locations and employers, and recording the demo were employed to improve the validity of findings in an interview study.

\emph{External Validity:}
The prototype is designed with specific components such as GPT-4 for code generation, esmini for simulation, and for generating Python code for ADS. However, the proposed pipeline is adaptable to other simulation environments and LLMs, as presented in Sec.~\ref{sec:evaluation_Models}. Additionally, the proposed pipeline can be used for other programming languages and functions. 

\section{Conclusion \& Future Work}
\label{sec:conlusion}

Rapid and efficient ADS software development and updates to expand functionality and ensure safety against newly identified hazardous events require novel approaches. LLM-based code generation is one approach; however, LLMs' reliability and accuracy pose challenges, especially for complex safety-related functions.
We designed and implemented a pipeline that provides Python code that represents the safety-related function description in the concept phase. This allows function and system developers to identify potential weaknesses and specify relevant safety requirements through automatically generated reports in natural language. Moreover, the generated code can serve as a formal notation for function description. The proposed pipeline is designed and implemented to be model-agnostic and flexible to incorporate new LLM models.


We initially identified and reported the limitations and insufficiencies of LLMs, either from the academic literature or through the design cycles of the pipeline. For instance, inaccuracy and incorrectness are well-known challenges that require human oversight. Studies report that correctness assessment, bug-fixing efforts, and prompt engineering difficulties pose significant challenges for human reviewers.
Hence, we designed and systematically evaluated a prototype consisting of an LLM-based code generator and an automated feedback mechanism for iterative safety evaluation and improvement.
This mechanism included an abstract simulation model, a set of safety acceptance criteria, and a report generator. The simulation model serves as an abstract and low-resource-intensive world model for the LLM to understand the potential effects of the generated code in the environment.
Safety acceptance criteria are implemented to automatically assess unsafe behaviours of the function, such as collisions, unintended activation, or harsh braking. These reports, derived from numerical logs, are readable by both the LLM for code improvement in the next iteration and by engineers for review. The generated report helps the LLM identify potential malfunctions or functional insufficiencies and iteratively improve the code.
Additionally, the pipeline enables a conversation between the LLM and the world model (i.e., the simulation), similar to a human-LLM conversation, which could serve as a mechanism to mitigate hallucinations in LLM-generated outputs.


Then, we evaluated the prototype in an industrial setup through a series of experiments using safety-related traffic scenarios.
The experimental results showed 
an average 9.2\% improvement in the number of satisfied test cases when using correction loop. Moreover, as reported in Fig.~\ref{fig:Results_all}, the first fully successful baseline was generated through correction by providing the simulation test report.

Multiple open-source models were examined in the pipeline for two functions and ranked based on their performance.
While most researchers rely on simplified, well-known tasks or benchmarks to assess LLM performance, our experiments reveal that these rankings can be misleading. Many models are specifically optimised to pass these known tasks and benchmarks (i.e., benchmark leakage) rather than to solve truly novel tasks.
As summarised in Fig.~\ref{fig:Results_all_models}, CodeGemma:7B and Mistral:7B were able to generate successful code for ACC (simpler function with potentially more leakage), but they failed to generate any successful code for CAEM (complex function with potentially less leakage). However, larger models such as CodeLlama:34B, DeepSeek (r1:32B and Coder:33B) were unable to deliver any successful code. Hence, GPT-4 is the only model that generated successful code for both ACC and CAEM.


Finally, the limitations and the designed pipeline were presented and discussed with experts from the automotive OEMs involved in ADS development through an interview study involving 11 experts.
Experts highlighted the tool's potential to reduce workload, save time, and aid in rapid prototyping, while also stressing the importance of human oversight to ensure the safety and correctness of the generated code. They also emphasised the necessity for suitable test scenarios and iterative updates to input requirements to address functional insufficiencies.

Future work should explore combining the proposed prototype with retrieval techniques~\cite{xu2021inide} to use certified legacy codes and integrate company-specific processes into the prompts, potentially improving quality and understandability. 
Additionally, employing formal methods to validate LLM-generated code, as proposed by P8, could improve software reliability. Finally, a comprehensive software tool qualification process is necessary to address the tool's impact, limitations, and reliability for specific use cases.

\section*{Acknowledgments}
This work has been partially supported by Sweden’s Innovation Agency (Vinnova, diarienummer: 2021-02585), and by the Wallenberg AI Autonomous Systems and Software Program (WASP) funded by the Knut and Alice Wallenberg Foundation.

\section*{Ethical Considerations}
\label{sec:Ethical Considerations}

As raised in discussions with the case companies, there is a risk that the code proposed by the LLM is the exact code owned by an individual, which is the intellectual property of that individual. The ethical, legal, and other relevant aspects of these and similar concerns require further study.
Moreover, These systems can be prompted in a way that generates harmful or biased outputs~\cite{chen2021codex}, which again emphasises the importance of human oversight and responsible usage. Other policies such as use case monitoring, restrictions, and limitation on usage rate, might also be used~\cite{chen2021codex} to reduce the risk.

\section*{Disclaimer}
The views and opinions expressed are those of the authors and do not necessarily reflect the official policy or position of Volvo Cars. The proposed methods, prompts, pipeline or the results generated by this work are only used in this study and not used in any engineering of production related projects. 
The proposed process is an initial step toward ADS software, and more activities need to be involved, such as code review by engineers, verification, validation, and confirmation review measures.

\end{document}